\documentclass[aps,prl,twocolumn,superscriptaddress]{revtex4-1}
\usepackage{color}
\usepackage{dcolumn}
\usepackage{graphicx}
\usepackage{amsmath}
\usepackage{amssymb}

\usepackage{amsmath,amsfonts,amssymb}

\usepackage{bbold}
\usepackage{dsfont}

\usepackage{soul}

\usepackage{fancyhdr}
\usepackage{multirow}
\usepackage{bm}
\usepackage{dsfont}
\usepackage{cancel}

\newcommand{\oH}{\hat{H}}
\newcommand{\oa}{\hat{a}}
\newcommand{\oad}{\hat{a}^{\dagger}}
\newcommand{\oGamma}{\hat{\Gamma}}
\newcommand{\oGammad}{\hat{\Gamma}^{\dagger}}
\newcommand{\orho}{\hat{\rho}}

\newcommand{\ket}[1]{|#1 \rangle}
\newcommand{\bra}[1]{\langle #1|}

\begin{document}

\title{Quantum critical regime in a quadratically-driven nonlinear photonic lattice}

\author{Riccardo Rota}
\affiliation{Institute of Physics, Ecole Polytechnique F\'ed\'erale de Lausanne (EPFL), CH-1015, Lausanne, Switzerland}
\email{riccardo.rota@epfl.ch}
\author{Fabrizio Minganti}
\affiliation{Laboratoire Mat\'eriaux et Ph\'enom\`enes Quantiques, Universit\'e Paris Diderot, CNRS-UMR 7162, 75013 Paris, France}
\affiliation{Theoretical Quantum Physics Laboratory, RIKEN Cluster for Pioneering Research, Wako-shi, Saitama 351-0198, Japan}
\author{Cristiano Ciuti}
\affiliation{Laboratoire Mat\'eriaux et Ph\'enom\`enes Quantiques, Universit\'e Paris Diderot, CNRS-UMR 7162, 75013 Paris, France}
\author{Vincenzo Savona}
\affiliation{Institute of Physics, Ecole Polytechnique F\'ed\'erale de Lausanne (EPFL), CH-1015, Lausanne, Switzerland}

\date{\today}

\begin{abstract}

We study an array of coupled optical cavities in presence of two-photon driving and dissipation. The system displays a critical behavior similar to that of a quantum Ising model at finite temperature. Using the corner-space renormalization method, we compute the steady-state properties of finite lattices of varying size, both in one- and two-dimensions. From a finite-size scaling of the average of the photon number parity, we highlight the emergence of a critical point in regimes of small dissipations, belonging to the quantum Ising universality class. For increasing photon loss rates, a departure from this universal behavior signals the onset of a quantum critical regime, where classical fluctuations induced by losses compete with long-range quantum correlations.

\end{abstract}

\maketitle

The emergence of critical phenomena in the non-equilibrium steady state (NESS) of open quantum systems \cite{CiutiRMP,Hartmann16,NohAngelakis16}, arising from the competition between the incoherent and coherent dynamics, is a topic that has gathered increasing attention in recent years, especially in view of the possible experimental realization of model systems using circuit QED \cite{Carmichael15,Fink17,Fitzpatrick17}, lattices of ultracold atoms \cite{Dimer07,Baumann10,Baumann11,Brennecke13}, or other advanced quantum platforms \cite{Muller20121,Bernien2017}. Several theoretical studies have highlighted the possibility of dissipative phase transitions in various many-body systems \cite{CiutiRMP,Hartmann16,NohAngelakis16,DallaTorre10,DallaTorre12,Lee13,Sieberer13,Sieberer14,Altman15,Carmichael15,Bartolo16,Mendoza16,Casteels16,Jin16,Maghrebi16,Rota17,Savona17,Casteels17,CasteelsFazio17,FossFeig17,Biondi17,Biella17,Vicentini18,Rota18}, possibly displaying novel universal properties \cite{Marino16}. In this regard, the question about the role played by quantum fluctuations in these critical phenomena is still a matter of debate.

Bosonic systems on a lattice have been the object of several studies, motivated by the analogy with the Hamiltonian Bose-Hubbard model and by the possibility to realize experimental studies using arrays of optical cavities with a third-order optical nonlinearity \cite{Biella17,Biondi17,Cao16,CasteelsFazio17,FossFeig17,Casteels17,LeBoite13,LeBoite14,Lebreuilly17,Wilson16,Lebreuilly18}. In recent years, the driven-dissipative Bose-Hubbard model in presence of a two-photon -- i.e. quadratic in the field -- driving term has been studied both theoretically \cite{Minganti2016,Bartolo16,Bartolo2017,GotoSciRep2016,GotoPRA16,Nigg17,Puri2017,PuriNJP17,Savona17,Munuz2018,BenitoPRA16,MingantiarXiv18_Spectral} and experimentally \cite{Leghtas15,Wang16}. This quadratically driven scheme has been in particular proposed as a possible realization of a noise-resilient quantum code, where photonic Schr\"odinger's cat states with even and odd photon number parity behave as interacting spin degrees of freedom \cite{Leghtas15,GotoSciRep2016,GotoPRA16,Nigg17,Puri2017}. This finding is suggestive of a possible scheme for realizing a photonic simulator of quantum spin models \cite{Aspuru-Guzik2012} and allows for a completely novel approach to the study of dissipative phase transitions. Indeed, while widely studied one-photon driving breaks the $U(1)$ symmetry of the Hamiltonian, two-photon driving preserves a $\mathbb{Z}_2$ symmetry which can be spontaneously broken \cite{Savona17,MingantiarXiv18_Spectral}, giving rise to a second order phase transition that is similar to that of the quantum transverse Ising model \cite{Dutta15}.

The analogy to a quantum spin model leads to expect that, in the limit of low losses, this model should display a quantum critical point (QCP) \cite{Sachdev2011} separating a symmetric and a broken symmetry phase, in the vicinity of which the system is characterized by extensive long-range entanglement. The classical fluctuations induced by small losses, may then induce a \textit{quantum critical regime} (QCR), i.e. a region in the phase diagram where classical fluctuations populate highly quantum correlated excitations and the spectral and dynamical properties of the system are characterized by a universal behavior which is today the object of intense investigations \cite{Kopp2005,Coleman2005,Sachdev2008,Gegenwart08,Sachdev2011a,Sachdev09}.

QCPs have been observed in magnetic insulators \cite{Bitko96,Ruegg08,Coldea10}, heavy-fermions metals \cite{Gegenwart08}, high-temperature superconductors \cite{Marel2003} and ultracold atoms \cite{Greiner02,Baumann10,Haller10,Zhang12}. Evidence of a QCR has recently been found in the magnetic phase diagram of cobalt niobate \cite{Kinross14}. The possibility to simulate the physics of the QCR on a driven-dissipative photonic platform would represent a unique opportunity to study this phenomenon in a controlled setting where all relevant correlations can be experimentally accessed \cite{Fitzpatrick17,Eichler14,Wang16}.

In this work, we will present a beyond mean-field investigation of a lattice of coupled optical resonators subject to a two-photon driving field and to one- and two-photon losses. We show that this system is characterized by a phase diagram which bears full analogy with that of the quantum transverse Ising model at finite temperature. For both 1D chains and 2D lattice, in the limit of small loss rates, we observe the emergence of a QCP belonging to the same universality class of the quantum transverse Ising model. At larger loss rates, a departure from the universal scaling characterizing the QCP suggests the occurrence of a QCR.

A lattice of coupled optical resonators, in presence of a Kerr nonlinearity and two-photon driving, is modeled by the quadratically-driven dissipative Bose-Hubbard model. In the reference frame rotating with half of the pump frequency ($\hbar = 1$), the Hamiltonian reads
\begin{eqnarray}
\oH &=& \sum_{j}-\Delta\oad_j\oa_j + \frac{U}{2}\hat{a}^{\dagger  2}_j\oa_j^2 + \frac{G}{2} \hat{a}^{\dagger  2}_j+ \frac{G^*}{2} \oa_j^2\nonumber\\
&-& \sum_{\langle j,j^\prime\rangle} \frac{J}{z} (\oad_j\oa_{j^\prime} + \oad_{j^\prime}\oa_{j}) \,.
\label{eq:hamiltonian}
\end{eqnarray}
Here $\oa_j$ is the photon destruction operator acting on the $j$-th site, $U$ is the energy associated to the Kerr nonlinearity, $G$ is the two-photon driving field amplitude, and $\Delta = \omega_p/2 - \omega_c$ is the detuning between half of the two-photon driving field frequency $\omega_p$ and the resonant cavity frequency $\omega_c$. The last term in the equation models the photon hopping between nearest-neighbor cavities $\langle j,j^\prime\rangle$, $J$ and $z$ being respectively the hopping strength and the number of nearest neighbors.

In presence of weak Markovian coupling to an environment, the system is described by the density matrix $\orho(t)$ which obeys the quantum master equation in Lindblad form
\begin{equation}
\frac{\partial \orho}{\partial t} = -i \left[\oH,\orho \right] + \sum_{j,k} \oGamma_{j,k}\orho\oGammad_{j,k}  - \frac{1}{2} \left\{\oGammad_{j,k}\oGamma_{j,k},\orho\right\} \,.
\label{eq:lindblad-me}
\end{equation}
Here, the jump operators $\oGamma_{j,1} = \sqrt{\gamma} \oa_j$ and $\oGamma_{j,2} = \sqrt{\eta} \oa^2_j$ induce respectively the one- and two-photon losses from the $j$-th cavity. In the limit of long time, the system evolves towards a non-equilibrium steady-state $\orho_{ss}$, which satisfies the condition $d\orho_{ss}/dt = 0$. 

The Hamiltonian \eqref{eq:hamiltonian} commutes with the parity of the total photon number operator $\hat{\Pi} = \exp(i \pi \sum_{j} \hat{a}^{\dagger}_j \hat{a}_j)$ and Eq. \eqref{eq:lindblad-me} is invariant under a global change of sign of the fields,  $\oa_j\rightarrow-\oa_j \  \forall j$, resulting in a $\mathbb{Z}_2$ symmetry of the system \cite{MingantiarXiv18_Spectral}. Thus, our model is expected to display a \textit{dissipative} phase transition associated to the spontaneous breaking of the $\mathbb{Z}_2$ symmetry, as suggested by a mean-field analysis \cite{Savona17}. In the absence of classical fluctuations, the phase transition may assume a \textit{quantum} nature.

In order to clarify the nature of the quantum phase transition in the present system, it is useful to briefly revisit the single-site problem. It was shown that the steady-state of the single Kerr resonator, for small detuning and for small loss rates, is well approximated by a statistical mixture of two Schr\"odinger cat states with opposite parity \cite{Minganti2016}, $\orho = p_{+} \ket{C_\alpha^{(+)}}\bra{C_\alpha^{(+)}} + p_{-} \ket{C_\alpha^{(-)}}\bra{C_\alpha^{(-)}}$. Here, $\ket{C_\alpha^{(\pm)}} = \left( \ket{\alpha} \pm \ket{-\alpha} \right)/\mathcal{N}_\alpha^{\pm}$, where $\ket{\alpha}$ is a coherent state and $\mathcal{N}_\alpha^{\pm} = \sqrt{2(1 \pm e^{-2|\alpha|^2})}$ is a normalization factor. The relevant local Hilbert space is therefore approximately spanned by two states only, as in a quantum $S=1/2$ spin system. The displacement $\alpha$ (which is generally complex, even for real-valued $G$) is determined uniquely by the values of the system parameters, and its amplitude can be varied by changing $G$ \cite{Note1}.

Hence, when the driving term is resonant with the minimum of the single particle energy band and the loss rates are small, the quadratically driven dissipative Bose-Hubbard model can be approximated with a spin model, where the even and odd cat states on each lattice site correspond respectively to the spin-up and spin-down states ($\ket{C_\alpha^{(+)}} \to \ket{\uparrow}$, $\ket{C_\alpha^{(-)}} \to \ket{\downarrow}$).
The Hamiltonian \eqref{eq:hamiltonian}, projected onto the cat state basis, is then expressed (up to a constant additive term) as
\begin{eqnarray}
& & \oH_{XY} = - \frac{\Delta |\alpha|^2}{2} A_- \sum_j \hat{\sigma}_{j}^{(z)} \label{eq:hamiltonianXY} \\
& & - \frac{J |\alpha|^2}{2z} \sum_{\langle j,j^\prime\rangle} \left[(A_+ + 2) \hat{\sigma}_{j}^{(x)} \hat{\sigma}_{j^\prime}^{(x)} + (A_+ - 2)  \hat{\sigma}_{j}^{(y)} \hat{\sigma}_{j^\prime}^{(y)}\right] \,, \nonumber
\end{eqnarray}
where $A_{\pm} = \tanh{|\alpha|^2} \pm (\tanh{|\alpha|^2})^{-1}$ and $\hat{\sigma}_{j}^{(l)}$ (with $l \in \{x,y,z\}$) are the Pauli matrices acting on the $j$-th spin \footnote{See Supplemental Material}.

\begin{figure}
	\begin{center}
		\includegraphics[width=0.48\textwidth]{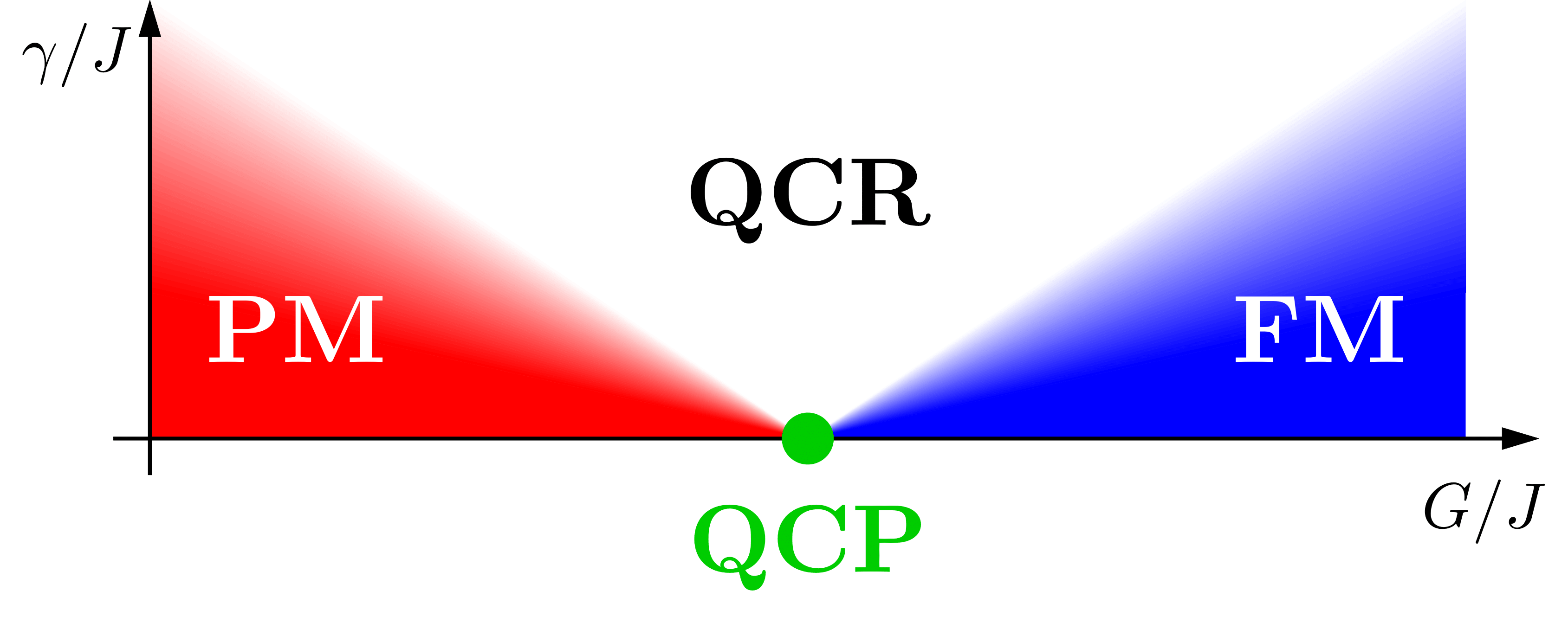}
	\end{center}
	\caption{A pictorial sketch of the phase diagram of the quadratically-driven dissipative Bose Hubbard model. For vanishing one-photon loss rate $\gamma$, when increasing the amplitude of the two-photon driving field $G$, the system undergoes a quantum phase transition. A Quantum Critical Point (QCP) marks the transition between a paramagnetic (PM) phase, where the NESS in the thermodynamic limit has a definite parity, and a ferromagnetic (FM) phase, where the $\mathbb{Z}_2$ symmetry of the system is spontaneously broken. For finite values of the loss rate $\gamma$, in the vicinity of the QCP, a Quantum Critical Regime (QCR) arises, where classical fluctuations act directly on the quantum-critical entangled state.}\label{fig:PhaseDiagram}
\end{figure}

\begin{figure*}
		\includegraphics[width=0.95\textwidth]{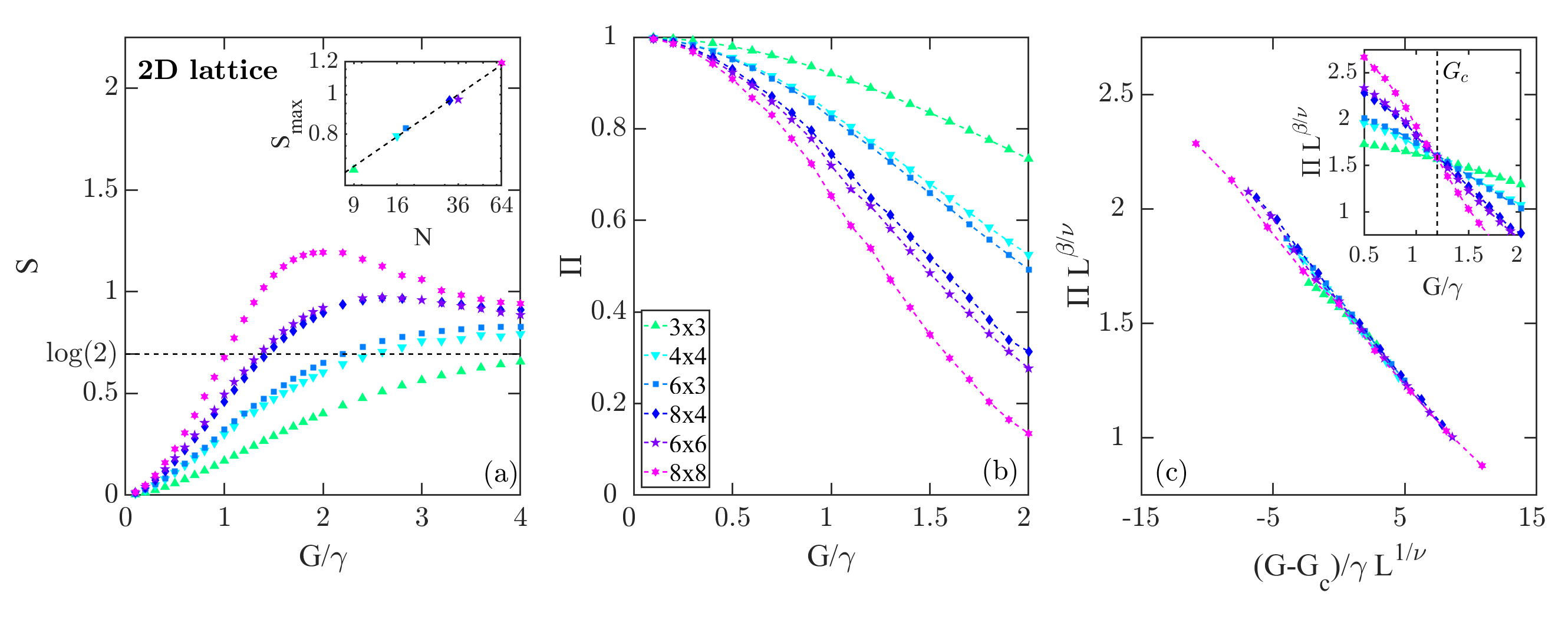}
	\caption{Computed values of the von-Neumann entropy $S$ (a) and of the parity $\Pi$ (b) as a function of $G/\gamma$ for the NESS of 2D lattices of different sizes. (c) Finite size scaling of the parity using the critical exponents of the quantum 2D transverse Ising model ($\beta = 0.32642$ and $\nu = 0.62997$). Inset: rescaled parity plotted vs $G/\gamma$. Parameters: $U/\gamma = 40$ and $J/\gamma = 20$.}\label{fig:Medium2D}
\end{figure*}

The Hamiltonian \eqref{eq:hamiltonianXY} is that of the quantum XY model, that is a generalization of the quantum transverse Ising model. It is characterized by a QCP both for 1D arrays and 2D lattices \cite{Dutta15}: the critical exponents of a $d$-dimensional quantum transverse Ising model are the same as a $(d+1)$-dimensional classical Ising model \cite{Sachdev2011}. In the present Bose-Hubbard model, a quantum phase transition can be induced by tuning the displacement $\alpha$ in Eq. \eqref{eq:hamiltonianXY} through the driving field $G$. In particular, in the limit of large $G/\gamma$ and thus of large $|\alpha|$, the XX-term dominates in Eq. \eqref{eq:hamiltonianXY}. This limit corresponds to a quantum Ising model at vanishing external field. Then, the $\mathbb{Z}_2$ symmetry is spontaneously broken and two equivalent states $\ket{\Psi_\pm}=\prod_j(\ket{\uparrow_j}\pm\ket{\downarrow_j})$ dominate the steady state of the system, in analogy with the ferromagnetic phase of the quantum Ising model. In this limit, the steady-state density matrix is approximated by a statistical mixture of the two dominant states with equal weights, and the von Neumann entropy correspondingly will approach $S=\log(2)$. Notice that two dominant quantum states in the large-$|\alpha|$ limit can also be expressed as $\ket{\Psi_\pm}=\ket{\pm\alpha,\ldots,\pm\alpha}$, thus recovering the mean-field picture of the broken symmetry \cite{GotoSciRep2016,GotoPRA16,Nigg17,Puri2017,Savona17}. When the value of $G/\gamma$ is reduced, the first term in Hamiltonian Eq. \eqref{eq:hamiltonianXY} becomes more relevant and plays the role of an external field along Z, eventually inducing a quantum phase transition. In the new phase, the dominant quantum state becomes $\ket{\Psi}=\prod_j\ket{\uparrow_j}$. This state has even parity and, for vanishing $|\alpha|$, coincides with the bosonic vacuum.

Unlike two-photon losses, one-photon losses $\hat{\Gamma}_{j,1}$ produce jumps between $\ket{C_\alpha^{(+)}}$ and $\ket{C_\alpha^{(-)}}$, and thus tend to mix states with an even number of photons to states with an odd one \cite{Note1}. The classical fluctuations induced by one-photon losses bear analogy to those present in a system at thermal equilibrium. Therefore, in the limit of small one-photon loss rate $\gamma\ll J$, and sufficiently close to the critical value $G_c$, the system is expected to be in a QCR, in analogy to the QCR occurring in systems at thermal equilibrium \cite{Sachdev2011,Sachdev2011a,Vojta03}. 
In this regime, the losses cause the NESS to become a statistical mixture of states, all characterized by strong and long-range quantum correlation. This behavior is illustrated in Fig. \ref{fig:PhaseDiagram}.

In order to study the emergence of a QCP in this system, we determine the steady-state density matrix of the full boson model described by the Lindblad Master equation, Eq. \eqref{eq:lindblad-me}, in the cases of 1D arrays and 2D lattices of different sizes, assuming periodic boundary conditions. We have posed $d\orho_{ss}/dt = 0$ in Eq. \eqref{eq:lindblad-me} and solved the resulting equation numerically. All cases studied here assume real values for $G$, $\gamma = \eta$ and $\Delta = - |J|$. 
The numerical solution of the steady state problem is obtained using the corner-space renormalization method \cite{FinazziPRL15}. 

We start our discussion considering the case of a 2D lattice. In Fig. \ref{fig:Medium2D} we show the results for the von Neumann entropy $S = -\textrm{Tr}(\hat{\rho}_{ss} \log \hat{\rho}_{ss})$ (panel (a)) and for the steady-state expectation value of the parity $\Pi = \textrm{Tr}\left(\hat{\rho}_{ss} \exp(i \pi \sum_{j} \hat{a}^{\dagger}_j \hat{a}_j)\right)$ (panel (b)) as a function of $G/\gamma$ for different size of the lattice, having fixed $U/\gamma = 40$ and $J/\gamma = 20$. We notice that in the two limiting regimes, we recover the expected phases: at small $G$, we find for any size of the lattice a pure steady state with positive parity, corresponding to the vacuum state. At large $G$ instead, the expectation value of the parity vanishes and the entropy approaches the value $S = \log(2)$, indicating that the steady state is a mixture of two equiprobable states with opposite parity. To investigate the emergence of the critical behavior, we study the scaling of the parity with the finite size $L=\sqrt{N}$ of the simulated two-dimensional lattice, according to the theory of Fisher and Barber \cite{Fisher72}. In this analysis, we use the critical exponents $\beta = 0.32642$ and $\nu = 0.62997$, respectively related to the magnetization and the correlation length for the quantum 2D transverse Ising model \cite{Vichi16}. In the inset of Fig. \ref{fig:Medium2D}-(c), we show the rescaled parity $\Pi L^{\beta/\nu}$ as a function of $G/\gamma$. The emergence of a critical point at $G_c/\gamma \simeq 1.2$ is indicated by the common crossing point of all the rescaled curves corresponding to different values of $L$. Moreover, when we plot the rescaled parity $\Pi L^{\beta/\nu}$ as a function of $(G - G_c)/\gamma \, L^{1/\nu}$ (main graph of Fig. \ref{fig:Medium2D}-(c)), the data collapse on a single universal curve, indicating that the phase transition belong to the universality class of the quantum 2D transverse Ising model. This result is consistent with the presence of long-range entanglement among photons, in spite of the nearest-neighbor interaction and the local dissipations.

The behavior of the entropy in the critical region brings insight into the role of classical fluctuations induced by the losses. For the largest lattices studied, $S$ is not monotonous with $G$, displaying instead a maximum, whose value scales as a power law of the number of sites $N$ of the lattice ($\max(S) \sim N^{\kappa}$ with a fitted value $\kappa = 0.29$: see inset fo Fig \ref{fig:Medium2D}-(a)). The sub-linear scaling of the entropy with the number of sites is a signature of correlations (classical and/or quantum) and, together with the emergence of a critical behavior belonging to the universality class of the quantum transverse Ising model, reveals the competing roles of quantum and classical fluctuations in the many-body system, thus supporting the QCR picture \cite{Sachdev2011}

\begin{figure}
		\includegraphics[width=0.48\textwidth]{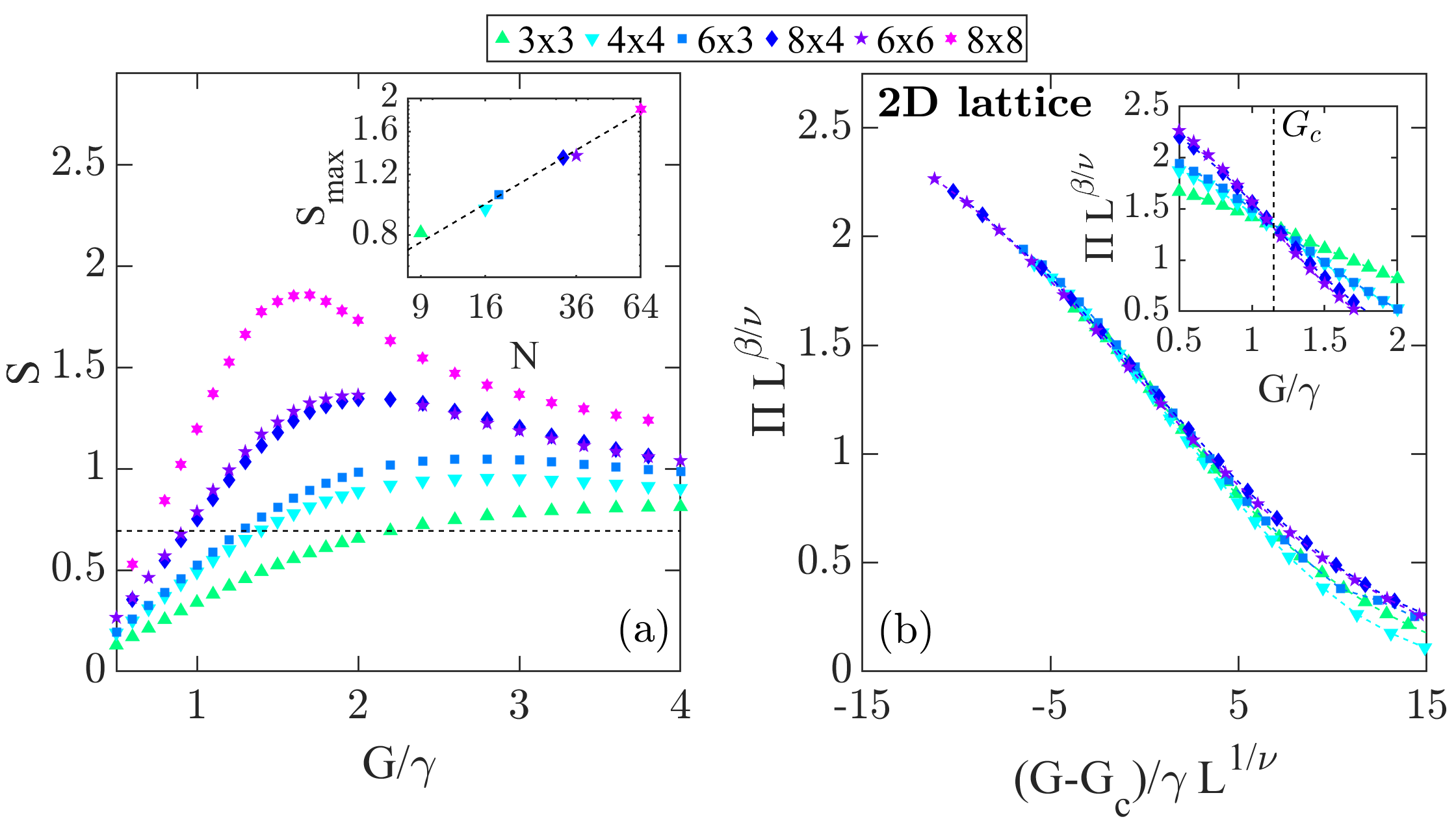}
	\caption{(a) von-Neumann entropy $S$ computed as a function of $G/\gamma$ for the NESS of 2D lattices of different sizes. (b) Finite size scaling analysis of the parity using the critical exponents of the quantum 2D transverse Ising model ($\beta = 0.32642$ and $\nu = 0.62997$). Inset: rescaled parity plotted vs $G/\gamma$. Parameters: $U/\gamma = 20$ and $J/\gamma = 10$.}\label{fig:Large2D}
\end{figure}

To better characterize the effect of classical fluctuations, we study the same two-dimensional lattice in a regime of larger dissipation ($U/\gamma = 20$ and $J/\gamma = 10$). The results are shown in Fig. \ref{fig:Large2D}. Panel (a) displays the entropy, whose peak value still increases as a power law of $N$, but with a larger fitted exponent $\kappa = 0.44$. The scaling analysis of the parity with the critical exponents of the 3-D Ising model, displayed in Fig. \ref{fig:Large2D}(b), shows that for lattice sizes up to $6 \times 6$, the rescaled data for  $\Pi L^{\beta/\nu}$ have a common crossing point at the same value of $G_c/\gamma \simeq 1.2$ found for smaller dissipation. However, the curves for $\Pi L^{\beta/\nu}$ as a function of $(G - G_c)/\gamma \, L^{1/\nu}$ display a sizeable departure from universality, particularly visible for $G>G_c$. 
We notice that, for this set of parameters, the convergence of the corner-space computed parity is extremely hard to assess in the $8 \times 8$ lattice, contrarily to the corresponding entropy data.

\begin{figure}
		\includegraphics[width=0.48\textwidth]{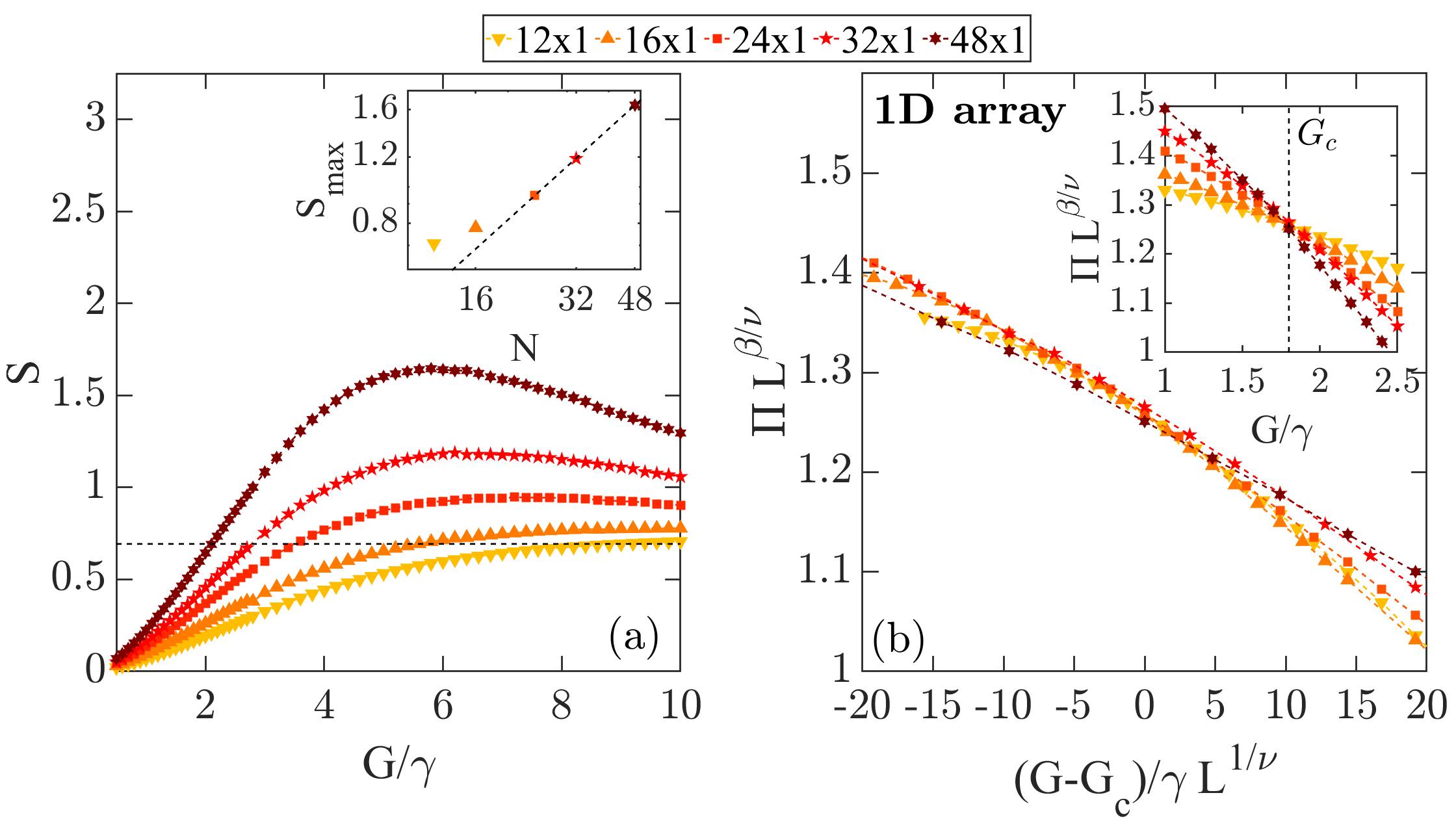}
	\caption{(a) von-Neumann entropy $S$ computed as a function of $G/\gamma$ for the NESS of 1D arrays of different lengths. (b) Finite size scaling analysis of the parity using the critical exponents of the quantum 1D transverse Ising model ($\beta = 0.125$ and $\nu = 1$). Inset: rescaled parity plotted vs $G/\gamma$. Parameters:  $U/\gamma = 100$ and $J/\gamma = 50$.}\label{fig:Small1D}
\end{figure}

A qualitatively similar behavior is observed when simulating a 1D array of cavities. In Fig. \ref{fig:Small1D}, we present the results obtained for the parameters $U/\gamma = 100$ and $\Delta/\gamma = 50$. In spite of the very small loss rate, the peak of the entropy still grows sub-linearly, but with a still larger exponent, i.e. $\max(S) \sim N^{\kappa}$ with a fitted value $\kappa = 0.80$. We perform a finite-size scaling analysis of the parity using the critical exponents of the quantum 1D transverse Ising model $\beta = 0.125$ and $\nu = 1$. As in the 2D case at larger loss rate, we find in this case a small departure from the universal scaling of the quantum phase transition. The plot of $\Pi L^{\beta/\nu}$ as a function of $G$ shows a common crossing point estimated at $G_c/\gamma \simeq 1.8$ (see inset of Fig. \ref{fig:Small1D}(b)), the rescaled plot of $\Pi L^{\beta/\nu}$ as a function of $(G - G_c)/\gamma \, L^{1/\nu}$ does not show a full collapse of the data at varying system size (main panel of Fig. \ref{fig:Small1D}(b)).

The picture that can be drawn from this analysis is that the driven-dissipative Bose-Hubbard model with two-photon driving behaves as a quantum simulator of an interacting spin model in presence of a transverse field. In the limit of vanishing loss rates, when varying the driving field amplitude $G$ across a critical value $G_c$, a quantum phase transition occurs from a paramagnetic phase to a ferromagnetic phase with broken $\mathbb{Z}_2$ symmetry. The critical exponents extracted from the numerical data correspond indeed to those of the quantum transverse Ising model and thus differ substantially from mean-field prediction, revealing the important role played by long-range entanglement arising among photons. Losses into the environment induce classical fluctuations that mimic those of a system at thermal equilibrium. For the 2D lattice at the smallest value of $\gamma$ that we have studied, the finite size of the lattices that we have considered is responsible for the suppression of excitations induced by classical fluctuations, and the scaling properties essentially reproduce the universal scaling expected at zero temperature. For the 2D case at larger loss rate and for the 1D case considered, the linear size of the system is large enough to enable classical fluctuations, and the scaling properties depart from the universal behavior, as one would expect for a quantum transverse Ising model at finite temperature. The distinctive feature of the present system is the fact that a quantum critical regime emerges for moderately small loss rates, that are within reach for example in circuit QED platforms \cite{Leghtas15}. In this limit, 
the system should display a quantum critical regime, characterized by a universal behavior of most dynamical properties, such as the excitation spectrum and the equilibration dynamics.

With the experimental feasibility of nonlinear photonic arrays \cite{Kuramochi14,Eichler14,Steger15,Wang16,Fitzpatrick17,Majumdar12,SavonaBadolatoArXiv,Baboux16,Jacqmin14}, the present result opens up the possibility to study the quantum critical regime on a highly controlled platform, where most dynamical, spectral and correlation properties are experimentally accessible. At the same time, the present system is highly versatile and could be generalized to the quantum simulation of several quantum spin models. We anticipate here that the study of the quadratically driven model with hopping rate $J<0$ -- an experimentally feasible regime \cite{antiferro} -- would enable the quantum simulation of antiferromagnetic spin models, possibly with frustrated lattice geometries. An extension to an $n$-photon driving field would instead introduce a $\mathbb{Z}_n$ symmetry, which was considered recently as a possible noise resilient quantum code \cite{Mirrahimi14} and has been recently realized in parametrically driven superconducting resonator \cite{Guo13,Svensson18}. The $\mathbb{Z}_n$ symmetry may open the way to the quantum simulation of spin $S>1/2$ models and of topological properties, as in the Haldane $S=1$ spin chain \cite{Affleck89}. Finally, a spatially non-homogeneous driving would naturally enable the study of many-body spin transport, while an engineered disorder would make it possible to study many-body localization and other properties of disordered interacting models.

We acknowledge enlightening discussions with Kenichi Komagata and Alessandro Vichi. This work was supported by the Swiss National Science Foundation through Project No. 200021\_162357 and by the European Research Council via Consolidator Grant CORPHO No. 616233.

\bibliographystyle{apsrev4-1}
\bibliography{CriticalQuadraticBIB}

\newpage
\onecolumngrid




\begin{center}
	\textbf{SUPPLEMENTAL MATERIAL}
\end{center}

	In this Supplemental Material, we firstly provide the detailed derivation of the Lindblad Master Equation for the effective spin system approximating the quadratically driven Bose-Hubbard Hamiltonian model. Then, we discuss an approximated expression for the displacement $\alpha$, needed to define the Schr\"odinger-cat states corresponding to the spin states, as a function of the physical parameters of the system.
	
	\section{Derivation of the Lindblad Master Equation for the spin model}
	
	In order to derive the Lindblad Master Equation of the spin model, let us start from the assumption of Ref. \cite{Minganti2016}, i.e. the steady-state density matrix of a single quadratically driven nonlinear resonator has two relevant eigenstates which are two cat-like states with opposite parity: $\ket{C_\alpha^{(\pm)}} = (\ket{\alpha} \pm \ket{-\alpha})/\mathcal{N}_\pm$. The normalization constants $\mathcal{N}_\pm$ are different for a finite value of $\alpha$: $\mathcal{N}_+ = \sqrt{\cosh(|\alpha|^2)/\exp(|\alpha|^2)}$ and $\mathcal{N}_- = \sqrt{\sinh(|\alpha|^2)/\exp(|\alpha|^2)}$.
	
	The annihilation operator $\hat{a}$ acts on these two states as:
	\begin{eqnarray}
	\hat{a} \ket{C_\alpha^{+}} & = & \frac{\alpha}{\mathcal{N}_+}(\ket{\alpha} - \ket{-\alpha}) = \alpha \frac{\mathcal{N}_-}{\mathcal{N}_+} \ket{C_\alpha^{-}} \ , \\
	\hat{a} \ket{C_\alpha^{-}} & = & \frac{\alpha}{\mathcal{N}_-}(\ket{\alpha} + \ket{-\alpha}) = \alpha \frac{\mathcal{N}_+}{\mathcal{N}_-} \ket{C_\alpha^{+}} \ .
	\end{eqnarray}
	Thus, if we truncate the local Hilbert space of a single resonator on the subspace spanned by these two states, $\hat{a}$ can be written in terms of the $2 \times 2$ matrix
	
	\begin{equation}\label{eq:Aspin}
	\hat{a} \simeq \alpha \left[ 
	\begin{array}{cc}
	0 & \frac{\mathcal{N}_+}{\mathcal{N}_-} \\
	\frac{\mathcal{N}_-}{\mathcal{N}_+} & 0 \\
	\end{array} 
	\right] = \frac{\alpha}{2} \left( B_x \hat{\sigma}^{(x)} - i B_y \hat{\sigma}^{(y)} \right) \ ,
	\end{equation}
	where in the last equality we have defined the real numbers
	\begin{eqnarray}
	B_x & = & \frac{\mathcal{N}_-}{\mathcal{N}_+} + \frac{\mathcal{N}_+}{\mathcal{N}_-} = \sqrt{\tanh(|\alpha|^2)} + \frac{1}{\sqrt{\tanh(|\alpha|^2)}} \ , \\
	B_y & = & \frac{\mathcal{N}_-}{\mathcal{N}_+} - \frac{\mathcal{N}_+}{\mathcal{N}_-} = \sqrt{\tanh(|\alpha|^2)} - \frac{1}{\sqrt{\tanh(|\alpha|^2)}} \ .
	\end{eqnarray}

	From the correspondence between the annihilation operator on a single cavity and the spin operators in Eq. \eqref{eq:Aspin}, we rewrite all the terms appearing in the sums of the Hamiltonian of the quadratically driven dissipative Bose-Hubbard model (Eq. (1) of the main text):
	\begin{itemize}
		\item{Detuning
			\begin{eqnarray}
			\Delta \hat{a}_j^\dagger \hat{a}_j & \simeq & \frac{\Delta |\alpha|^2}{4} \left(  B_x \hat{\sigma}_j^{(x)} + i B_y \hat{\sigma}_j^{(y)} \right)  \left(  B_x \hat{\sigma}_j^{(x)} - i B_y \hat{\sigma}_j^{(y)} \right) \nonumber \\
			& = & \frac{\Delta |\alpha|^2}{4} \left( B_x^2 \underbrace{\hat{\sigma}_j^{(x)} \hat{\sigma}_j^{(x)}}_{\mathds{1}}  + i B_x B_y \underbrace{(\hat{\sigma}_j^{(y)} \hat{\sigma}_j^{(x)} - \hat{\sigma}_j^{(x)} \hat{\sigma}_j^{(y)})}_{- 2 i \hat{\sigma}_j^{(z)}} + B_y^2 \underbrace{\hat{\sigma}_j^{(y)} \hat{\sigma}_j^{(y)}}_{\mathds{1}} \right) \nonumber \\
			& = & \frac{\Delta |\alpha|^2}{4} (B_x^2 + B_y^2) \mathds{1} + \frac{\Delta |\alpha|^2}{2} B_x B_y \hat{\sigma}_j^{(z)}
			\end{eqnarray}
		}
		\item{Two-photon driving
			\begin{eqnarray}
			\frac{G}{2} \hat{a}_j^{\dagger 2} + \frac{G^*}{2} \hat{a}_j^{2} & \simeq & \frac{G \alpha^{* 2}}{8} \left(  B_x \hat{\sigma}_j^{(x)} + i B_y \hat{\sigma}_j^{(y)} \right)^2 + \frac{G^* \alpha^2}{8} \left(  B_x \hat{\sigma}_j^{(x)} - i B_y \hat{\sigma}_j^{(y)} \right)^2 \nonumber \\
			& = & \frac{G\alpha^{* 2} + G^* \alpha^2}{8} (B_x^2 \underbrace{\hat{\sigma}_j^{(x)} \hat{\sigma}_j^{(x)}}_{\mathds{1}} - B_y^2 \underbrace{\hat{\sigma}_j^{(y)} \hat{\sigma}_j^{(y)}}_{\mathds{1}} ) + i \frac{G\alpha^{* 2} - G^* \alpha^2}{8} B_x B_y \underbrace{(\hat{\sigma}_j^{(y)} \hat{\sigma}_j^{(x)} + \hat{\sigma}_j^{(x)} \hat{\sigma}_j^{(y)})}_{0} \nonumber \\
			& = & \frac{G\alpha^{* 2} + G^* \alpha^2}{8} \left(B_x^2 - B_y^2 \right) \mathds{1}
			\end{eqnarray}
		}
		\item{Nonlinearity
			\begin{eqnarray}
			\frac{U}{2} \hat{a}_j^{\dagger 2} \hat{a}_j^{2} & \simeq & \frac{U |\alpha|^4}{8} \left(  B_x \hat{\sigma}_j^{(x)} + i B_y \hat{\sigma}_j^{(y)} \right)^2 \left(  B_x \hat{\sigma}_j^{(x)} - i B_y \hat{\sigma}_j^{(y)} \right)^2 \nonumber \\
			& = & \frac{U |\alpha|^4}{8} (B_x^2-B_y^2)^2 \mathds{1}
			\end{eqnarray}
		}
		\item{Hopping
			\begin{eqnarray}
			\frac{J}{z} (\hat{a}_j^{\dagger} \hat{a}_{j'} + \hat{a}_{j'}^{\dagger} \hat{a}_j) & \simeq & \frac{J|\alpha|^2}{4z} \left[ \left(  B_x \hat{\sigma}_j^{(x)} + i B_y \hat{\sigma}_j^{(y)} \right)  \left(  B_x \hat{\sigma}_{j'}^{(x)} - i B_y \hat{\sigma}_{j'}^{(y)} \right) + \left(  B_x \hat{\sigma}_{j'}^{(x)} + i B_y \hat{\sigma}_{j'}^{(y)} \right)  \left(  B_x \hat{\sigma}_j^{(x)} - i B_y \hat{\sigma}_j^{(y)} \right) \right] \nonumber \\
			& = & \frac{J|\alpha|^2}{4z} \left( B_x^2 \hat{\sigma}_{j}^{(x)} \hat{\sigma}_{j'}^{(x)} - \cancel{i B_x B_y  \hat{\sigma}_{j}^{(x)} \hat{\sigma}_{j'}^{(y)}} + \cancel{i B_x B_y  \hat{\sigma}_{j}^{(y)} \hat{\sigma}_{j'}^{(x)}} +  B_y^2 \hat{\sigma}_{j}^{(y)} \hat{\sigma}_{j'}^{(y)} \right.  \nonumber \\
			& & + \left. B_x^2 \hat{\sigma}_{j'}^{(x)} \hat{\sigma}_{j}^{(x)} - \cancel{i B_x B_y  \hat{\sigma}_{j'}^{(x)} \hat{\sigma}_{j}^{(y)}} + \cancel{i B_x B_y  \hat{\sigma}_{j'}^{(y)} \hat{\sigma}_{j}^{(x)}} +  B_y^2 \hat{\sigma}_{j'}^{(y)} \hat{\sigma}_{j}^{(y)} \right)  \nonumber \\
			& = & \frac{J|\alpha|^2}{2z} \left( B_x^2 \hat{\sigma}_{j}^{(x)} \hat{\sigma}_{j'}^{(x)} + B_y^2 \hat{\sigma}_{j}^{(y)} \hat{\sigma}_{j'}^{(y)} \right)
			\end{eqnarray}
		}
	\end{itemize}
	
	If we insert all these term in the full Hamiltonian and we drop out the terms proportional to the identity matrix, we get the Hamiltonian of the quantum XY model in a transverse filed
	\begin{equation}
	\hat{H}_{XY} = \sum_j -\frac{\Delta |\alpha|^2}{2} B_x B_y \hat{\sigma}_j^{(z)} - \sum_{\langle j, j' \rangle}  \frac{J|\alpha|^2}{4z} \left( B_x^2 \hat{\sigma}_{j}^{(x)} \hat{\sigma}_{j'}^{(x)} + B_y^2 \hat{\sigma}_{j}^{(y)} \hat{\sigma}_{j'}^{(y)} \right) \ .
	\end{equation}
	The last equation is equivalent to Eq. (3) of the main text, having noticed that
	\begin{eqnarray}
	B_x B_y & = & \left( \sqrt{\tanh(|\alpha|^2)} + \frac{1}{\sqrt{\tanh(|\alpha|^2)}}\right) \left( \sqrt{\tanh(|\alpha|^2)} - \frac{1}{\sqrt{\tanh(|\alpha|^2)}} \right) = \tanh(|\alpha|^2) - \frac{1}{\tanh(|\alpha|^2)} = A_- \nonumber \\
	B_x^2 & = & \left( \sqrt{\tanh(|\alpha|^2)} + \frac{1}{\sqrt{\tanh(|\alpha|^2)}}\right)^2 = \tanh(|\alpha|^2) + \frac{1}{\tanh(|\alpha|^2)} + 2 = A_+ + 2 \nonumber \\
	B_y^2 & = & \left( \sqrt{\tanh(|\alpha|^2)} - \frac{1}{\sqrt{\tanh(|\alpha|^2)}}\right)^2 = \tanh(|\alpha|^2) + \frac{1}{\tanh(|\alpha|^2)} - 2 = A_+ - 2 \nonumber 
	\end{eqnarray}
	with the definition $A_{\pm} = \tanh{|\alpha|^2} \pm (\tanh{|\alpha|^2})^{-1}$, given in the main text.
	
	Using the approximation of $\hat{a}_j$ given in Eq. \eqref{eq:Aspin}, we rewrite the jump operators appearing in the Lindblad master equation (Eq. (2) of the main text) within the formalism of the spin model:
	
	\begin{itemize}
		\item{
			One-photon losses:
			\begin{eqnarray}
			\hat{\Gamma}_{j,1} =\sqrt{\gamma} \hat{a}_j & \simeq & \frac{\sqrt{\gamma} \alpha}{2} \left( B_x \hat{\sigma}^{(x)}_j - i B_y \hat{\sigma}^{(y)}_j \right) \label{eq:onephotonlosses}
			\end{eqnarray}
		}
		\item{
			Two-photon losses:
			\begin{eqnarray}
			\hat{\Gamma}_{j,2} =\sqrt{\eta} \hat{a}^2_j & \simeq & \frac{\sqrt{\eta} \alpha^2}{4} \left( B_x \hat{\sigma}^{(x)}_j - i B_y \hat{\sigma}^{(y)}_j \right)^2 = \frac{\sqrt{\eta} \alpha^2}{4} \left(B_x^2 - B_y^2 \right) \mathds{1}
			\end{eqnarray}
		}
		
		Since $\hat{\Gamma}_{j,2}$ is approximated by a term proportional to the identity, the term
		\begin{equation}
		\oGamma_{j,2}\orho\oGammad_{j,2}  - \frac{1}{2} \left\{\oGammad_{j,2}\oGamma_{j,2},\orho\right\} \simeq 0
		\end{equation}
		for all values of $j$. For this reason, two-photon losses do not affect the dynamics of the effective spin model. The Lindblad master equation for the density matrix $\orho_{sp}$ in the spin model thus depends only on the jump operators $\oGamma_{j,1}$, approximated as in Eq. \eqref{eq:onephotonlosses}, and finally reads:
		\begin{eqnarray}
		\frac{\partial \orho_{sp}}{\partial t} & = & -i \left[\oH_{XY},\orho_{sp} \right] \nonumber  \\
		& + & \frac{\gamma |\alpha|^2}{4}\sum_{j} \left( B_x \hat{\sigma}^{(x)}_j - i B_y \hat{\sigma}^{(y)}_j \right) \orho_{sp} \left( B_x \hat{\sigma}^{(x)}_j + i B_y \hat{\sigma}^{(y)}_j \right) - \frac{1}{2} \left\{(B_x^2 + B_y^2) \mathds{1} + B_x B_y \hat{\sigma}_j^{(z)}, \orho_{sp}\right\} \ . \label{eq:MasterSpin}
		\end{eqnarray}
		This equation notably simplifies in the regimes of $\alpha \ll 1$ (small pumping) and $\alpha \gg 1$ (large pumping). In the first case, $\oGamma_{j,1}$ would become proportional to the spin raising operator $\hat{\sigma}^+_j$ and Eq. \eqref{eq:MasterSpin} would approximate as 
		\begin{equation}
		\frac{\partial \orho_{sp}}{\partial t} = -i \left[\oH_{XY},\orho_{sp} \right] + \frac{\gamma}{4}\sum_{j} \hat{\sigma}^{+}_j \orho_{sp} \hat{\sigma}^{-}_j - \frac{1}{2} \left\{\hat{\sigma}^{-}_j \hat{\sigma}^{+}_j , \orho_{sp}\right\} \ \ \ \ \ \ \ \ \ \textrm{for} \  \alpha \ll 1
		\end{equation}
		In the latter case, $\oGamma_{j,1}$ would become proportional to $\hat{\sigma}^{(x)}_j$ and Eq. \eqref{eq:MasterSpin} would approximate as 
		\begin{equation}
		\frac{\partial \orho_{sp}}{\partial t} = -i \left[\oH_{XY},\orho_{sp} \right] + \gamma |\alpha|^2\sum_{j} \hat{\sigma}^{(x)}_j \orho_{sp} \hat{\sigma}^{(x)}_j - \frac{1}{2} \left\{ \mathds{1} , \orho_{sp}\right\} \ \ \ \ \ \ \ \ \ \textrm{for} \  \alpha \gg 1
		\end{equation}

	\end{itemize}

	\section{Tunability of the displacement $\pmb{\alpha}$ with physical parameters}\label{sec:semiclassical}
	
	In this section, we follow a semiclassical approach to study a single quadratically-driven cavity, in order to find an approximated expression for the displacement $\alpha$ as a function of the parameters of the physical system. Although this approximation may not be reliable in all the range of parameters considered in the work, it gives us insight on the tunability of $\alpha$ with the amplitude of the two-photon driving $G$. 
	
	In a semiclassical regime, the displacement can be understood as a mean field $\alpha(t) = \textrm{Tr}\left[ \orho(t) \oa \right]$. From the Lindblad master equation for $\orho(t)$ (Eq. (2) of the main text, with $J=0$), one can derive the equation of motion of $\alpha(t)$:
	\begin{equation}
	\dot{\alpha} = i \left[ -\Delta \alpha + G \alpha^* + U |\alpha|^2 \alpha \right] - \frac{\gamma}{2} \alpha - \eta |\alpha|^2 \alpha
	\end{equation}
	
	The condition for the steady state ($\dot{\alpha} = 0$) leads to the following nonlinear equation for $\alpha$:
	\begin{equation}
	- \left(\Delta - i \frac{\gamma}{2} \right) \alpha + G \alpha^* + (U - i \eta) |\alpha|^2 \alpha = 0 \ .
	\end{equation}
	
	This equation has a trivial solution $\alpha = 0$. Moreover, since the detuning and the loss rates has to be small in order to approximate the photonic system with the spin model, one can drop the first term proportional to $\alpha$ and recover a simple expression for the second solution $\alpha = |\alpha| e^{i \phi}$, with:
	
	\begin{equation}
	|\alpha|^2 = \frac{G}{\sqrt{U^2 + \eta^2}} \ \ \ \ \ \ \ \ \phi = \frac{i}{2} \ln \left(\frac{i \eta - U}{\sqrt{U^2 + \eta^2}}\right) \ .
	\end{equation}

\end{document}